\documentclass[12pt]{revtex4}
\usepackage{bm}

\newcommand{\av}{{\boldsymbol A}}
\newcommand{\ev}{{\boldsymbol E}}
\newcommand{\jv}{{\boldsymbol j}}
\newcommand{\uv}{{\boldsymbol u}}
\newcommand{\grad}{{\boldsymbol \nabla}}

\begin{document}

\title{On vector potential of the Coulomb gauge}
\author{Valery P Dmitriyev}
\address{Lomonosov University, P.O. Box 160, Moscow 117574, Russia}
\email{dmitr@cc.nifhi.ac.ru}
%\date{20 August 2003}

\begin {abstract} The question of an instantaneous action (A M Stewart 2003 Eur. J. Phys.
\textbf{24} 519) can be approached in a systematic way applying
the Helmhotz vector decomposition theorem to a two-parameter
Lorenz-like gauge. We thus show that only the scalar potential may
act instantaneously.
\end {abstract}

%\submitto{EJP}


\maketitle
\section{Introduction}

The role of the gauge condition in classical electrodynamics was
recently highlighted \cite{Jackson}. This is because of probable
asymmetry between different gauges. The distinct feature of the
Coulomb gauge is that it implies an instantaneous action of the
scalar potential \cite{Stewart, Drury, Chubykalo}. The question of
simultaneous co-existence of instantaneous and retarded
interactions is mostly debated \cite{Jackson1}. The paper
\cite{Stewart} concludes that `the vector decomposition theorem of
Helmholtz leads to a form of the vector potential of the Coulomb
gauge that, like the scalar potential, is instantaneous'. This
conclusion was arrived at considering the retarded integrals for
electrodynamic potentials. Constructing within the same theorem
wave equations the author \cite{Drury} finds that `the scalar
potential propagates at infinite speed while the vector potential
propagates at speed $c$ in free space'. In order to resolve the
discrepancy between \cite{Stewart} and \cite{Drury} the latter
technique will be developed below in a more systematic way.

Recently the two-parameter generalization of the Lorenz gauge was
considered \cite{Jackson, Chubykalo}:
\begin{equation}
\grad \cdot {\boldsymbol A} + \frac{c}{c_{\rm g}^2} \frac{\partial
\varphi}{\partial t} = 0 \,, \label{0.1}
\end{equation}
where $c_{\rm g}$ is a constant that may differ from $c$ . We will
construct wave equations applying  the vector decomposition
theorem to Maxwell's equations with (\ref{0.1}). Thus simultaneous
co-existence of instantaneous and retarded actions will be
substantiated.

\section{Maxwell's equations in the Kelvin-Helmholtz
representation}

Maxwell's equations in terms of electromagnetic potentials ${\bf
A}$ and $\varphi$ read as
\begin{eqnarray}
\frac{1}{c}\frac{\partial \av}{\partial t} + \ev + \grad
\varphi = 0 \label{1.1}\\
\frac{\partial \ev}{\partial t} - c\grad \times (\grad \times
\av) + 4\pi \jv = 0 \label{1.2}\\
 \grad \cdot \ev = 4\pi
\rho \,. \label{1.3}
\end{eqnarray}
The Helmholtz theorem says that a vector field $\uv$ that vanishes
at infinity can be expanded into a sum of its solenoidal $\uv_{\rm
r}$ and irrotational $\uv_{\rm g}$ components. We have for the
electric field:
\begin{equation}
\ev = \ev_{\rm r} + \ev_{\rm g} \,, \label{1.4}
\end{equation}
where
\begin{eqnarray}
 \grad \cdot \ev_{\rm r} &=& 0 \label{1.5}\\
 \grad \times \ev_{\rm g} &=& 0 \,. \label{1.6}
\end{eqnarray}
The similar expansion for the vector potential can be written as
\begin{equation}
\av = \av_{\rm r} + \frac{c}{c_{\rm g}} \av_{\rm g} \,,
\label{1.7}
\end{equation}
where
\begin{eqnarray}
\grad
\cdot \av_{\rm r} &=& 0 \label{1.8}\\
\grad \times \av_{\rm g} &=& 0 \, . \label{1.9}
\end{eqnarray}
If we substitute Eqs.~(\ref{1.4}) and (\ref{1.7}) into
Eq.~(\ref{1.1}), we obtain
\begin{equation}
\frac{1}{c}\frac{\partial \av_{\rm r}}{\partial t} + \ev_{\rm r} +
\frac{1}{c_{\rm g}}\frac{\partial \av_{\rm g}}{\partial t} +
\ev_{\rm g} + \grad \varphi = 0 \,. \label{1.10}
\end{equation}
By taking the curl of Eq.~(\ref{1.10}), we obtain using
Eqs.~(\ref{1.6}) and (\ref{1.9})
\begin{equation}
 \grad \times \biggl(
{\frac{1}{c} \frac{\partial \av_{\rm r}}{\partial t} + \ev_{\rm
r}} \biggr) = 0 \,. \label{1.11}
\end{equation}
On the other hand, from Eqs.~(\ref{1.5}) and (\ref{1.8}), we have
\begin{equation}
 \grad \cdot \biggl(
{\frac{1}{c}\frac{\partial \av_{\rm r}}{\partial t} + \ev_{\rm r}}
\biggr) = 0 \,. \label{1.12}
\end{equation}
If the divergence and curl of a field are zero everywhere, then
that field must vanish. Hence, Eqs.~(\ref{1.11}) and (\ref{1.12})
imply that
\begin{equation}
\frac{1}{c}\frac{\partial \av_{\rm r}}{\partial t} + \ev_{\rm r} =
0 \,. \label{1.13}
\end{equation}
We subtract Eq.~(\ref{1.13}) from Eq.~(\ref{1.10}) and obtain
\begin{equation}
 \frac{1}{c_{\rm
g}}\frac{\partial \av_{\rm g}}{\partial t} + \ev_{\rm g} + \grad
\varphi = 0 \,. \label{1.14}
\end{equation}
Similarly, if we express the current density as
\begin{equation}
\jv = \jv_{\rm r} + \jv_{\rm g}\, , \label{1.15}
\end{equation} where
\begin{eqnarray}
 \grad \cdot \jv_{\rm r} &=& 0 \label{1.16}\\
\grad \times \jv_{\rm g} &=& 0 \,, \label{1.17}
\end{eqnarray}
Eq.~(\ref{1.2}) can be written as two equations
\begin{equation}
 \frac{\partial \ev_{\rm
r}}{\partial t} - c \grad \times (\grad \times \av_{\rm r}) + 4\pi
\jv_{\rm r} = 0 \label{1.18}
\end{equation}
\begin{equation}
\frac{\partial \ev_{\rm g}}{\partial t} + 4\pi \jv_{\rm g} = 0 \,.
\label{1.19}
\end{equation}
{}From Eqs.~(\ref{1.4}) and (\ref{1.5}), Eq.~(\ref{1.3}) can be
expressed as
\begin{equation}
\grad \cdot \ev_{\rm g} = 4\pi \rho \,. \label{1.20}
\end{equation}

\section{Wave equations for the two-speed extension of electrodynamics}

We will derive from Eqs.~(\ref{1.13}), (\ref{1.14}), (\ref{1.18}),
(\ref{1.19}), and (\ref{1.20}) the wave equations for the
solenoidal (transverse) and irrotational (longitudinal) components
of the fields. In what follows we will use the general vector
relation
\begin{equation}
\grad ( {\grad \cdot \uv}) = \grad^2 \uv + \grad \times (\grad
\times \uv)\,. \label{2.1}
\end{equation}
The wave equation for $\av_{\rm r}$ can now be found. We
differentiate Eq.~(\ref{1.13}) with respect to time:
\begin{equation}
\frac{1}{c}\frac{\partial^2 \av_{\rm r}}{\partial t^2} +
\frac{\partial \ev_{\rm r}}{\partial t} = 0 \,. \label{2.2}
\end{equation}
We next substitute Eq.~(\ref{1.18}) into Eq.~(\ref{2.2}) and use
Eqs.~(\ref{2.1}) and (\ref{1.8}) to obtain
\begin{equation}
 \frac{\partial^2 \av_{\rm r}}{\partial t^2} - c^2 \grad^2
\av_{\rm r} = 4\pi c\jv_{\rm r} \,. \label{2.3}
\end{equation}

The wave equation for $\ev_{\rm r}$ can be found as follows. We
differentiate Eq.~(\ref{1.18}) with respect to time:
\begin{equation}
\frac{\partial^2 \ev_{\rm r}}{\partial t^2} - c\grad \times (\grad
\times \frac{\partial \av_{\rm r}}{\partial t}) + 4\pi
\frac{\partial \jv_{\rm r}}{\partial t} = 0 \,, \label{2.4}
\end{equation}
and substitute Eq.~(\ref{1.13}) into Eq.~(\ref{2.4}). By using
Eqs.~(\ref{2.1}) and (\ref{1.5}), we obtain
\begin{equation}
 \frac{\partial^2\ev_{\rm r}}{\partial t^2}
 - c^2 \grad^2 \ev_{\rm r} = -4\pi
\frac{\partial \jv_{\rm r}}{\partial t} \,. \label{2.5}
\end{equation}
In the absence of the electric current, Eqs.~(\ref{2.3}) and
(\ref{2.5}) are wave equations for the solenoidal fields $\av_{\rm
r}$ and $\ev_{\rm r}$.

To find wave equations for the irrotational fields, we need a
gauge relation. Substituting (\ref{1.7}) into Eq.~(\ref{0.1}) we
get the longitudinal gauge
\begin{equation}
 \grad \cdot \av_{\rm g} + \frac{1}{{c}_{\rm g}} \frac{\partial
\varphi}{\partial t} = 0 \, . \label{2.6}
\end{equation}
The solenoidal part of the vector potential automatically
satisfies the Coulomb gauge, Eq.~(\ref{1.8}). The wave equation
for $\av_{\rm g}$ can be found as follows. We first differentiate
Eq.~(\ref{1.14}) with respect to time:
\begin{equation}
\frac{1}{c_{\rm g}}\frac{\partial^2\av_{\rm g}}{\partial t^2} +
\frac{\partial \ev_{\rm g}}{\partial t} + \frac{\partial \grad
\varphi}{\partial t} = 0 \,. \label{2.7}
\end{equation}
We then take the gradient of Eq.~(\ref{2.6}),
\begin{equation}
\grad ( {\grad \cdot \av_{\rm g} }) + \frac{1}{c_{\rm g}} \grad
\frac{\partial \varphi}{\partial t} = 0 \,, \label{2.8}
\end{equation}
and combine Eqs.~(\ref{2.7}), (\ref{2.8}) and (\ref{1.19}). If we
use Eqs.~(\ref{2.1}) and (\ref{1.9}), we obtain
\begin{equation}
\frac{\partial^2 \av_{\rm g}}{\partial t^2} - c_{\rm g}^2 \grad^2
\av_{\rm g} = 4\pi c_{\rm g}\jv_{\rm g} \,. \label{2.9}
\end{equation}

Next, we will find the wave equation for $\varphi$. We take the
divergence of Eq.~(\ref{1.14}),
\begin{equation}
\frac{1}{c_{\rm g}}\frac{\partial \grad \cdot \av_{\rm
g}}{\partial t} + \grad \cdot \ev_{\rm g} + \grad^2 \varphi = 0
\,, \label{2.10}
\end{equation}
and combine Eqs.~(\ref{2.10}), (\ref{2.6}), and (\ref{1.20}):
\begin{equation}
 \frac{\partial^2\varphi}{\partial t^2}
 - c_{\rm g}^2 \grad^2 \varphi = 4\pi c_{\rm g}^{\rm 2} \rho
\, . \label{2.11}
\end{equation}
Equations~(\ref{2.11}) and (\ref{2.9}) give wave equations for
$\varphi$ and $\av_{\rm g}$.

We may try to find a wave equation for $\ev_{\rm g}$ using
Eq.~(\ref{1.14}) in Eqs.~(\ref{2.11}) and (\ref{2.9}). However, in
the absence of the charge, we have from Eq.~(\ref{1.20})
\begin{equation}
 \grad \cdot \ev_{\rm g} = 0 \,. \label{2.15}
\end{equation}
Hence, by Eqs.~(\ref{2.15}) and (\ref{1.6}), we have
\begin{equation}
\ev_{\rm g}=0\,. \label{2.16}
\end{equation}

We see that Maxwell's equations (\ref{1.1})--(\ref{1.3}) with the
longitudinal gauge (\ref{2.6}) imply that the solenoidal and
irrotational components of the fields propagate with different
velocities. The solenoidal components $\av_{\rm r}$ and $\ev_{\rm
r}$ propagate with the speed $c$ of light, and the irrotational
component $\av_{\rm g}$ of the magnetic vector potential and the
scalar potential $\varphi$ propagate with the speed $c_{\rm g}$.

\section{Single-parameter electrodynamics}
In reality, electrodynamics has only one parameter, the speed of
light, $c$.  Then, to construct from the above the classical
theory, we have to choose among two variants: two waves with equal
speeds or a single wave. If we let
\begin{equation}
 c_{\rm g} = c, \label{3.1}
\end{equation}
the two-parameter form (\ref{0.1}) becomes the familiar Lorenz
gauge
\begin{equation}
\grad \cdot \av + \frac{1}{c} \frac{\partial \varphi}{\partial t}
= 0 \, . \label{3.2}
\end{equation}
Another possible choice is
\begin{equation}
 c_{\rm g} \gg c \, . \label{3.3}
\end{equation}
The condition  (\ref{3.3}) turns Eq.~(\ref{0.1}) into the Coulomb
gauge
\begin{equation}
\grad \cdot\av = 0 \,. \label{3.4}
\end{equation}
Substituting
\begin{equation}
 c_{\rm g} = \infty \, \label{3.5}
\end{equation}
into the dynamic equation (\ref{2.11}), we get
\begin{equation}
 - \grad^2 \varphi = 4\pi \rho \, . \label{3.6}
\end{equation}
Validity of Eq.~(\ref{3.6}) for the case when $\varphi$ and $\rho$
may be functions of time $t$ means that the scalar potential
$\varphi$ acts instantaneously.

Substituting (\ref{3.5}) into Eq.~(\ref{2.6}) we get for the
irrotational part of the vector potential:
\begin{equation}
\grad \cdot\av_{\rm g} = 0\,. \label{3.7}
\end{equation}
Insofar as the divergence (\ref{3.7}) and the curl (\ref{1.9}) of
$\av_{\rm g}$ are vanishing, we have
\begin{equation}
\av_{\rm g} = 0\,. \label{3.8}
\end{equation}
So, on the first sight by (\ref{3.5}) irrotational component
$\av_{\rm g}$ of the vector potential propagates instantaneously.
However, according to relation (\ref{3.8}), with (\ref{3.5})
$\av_{\rm g}$ vanishes.

Putting (\ref{3.8}) into (\ref{1.7}) we get
\begin{equation}
\av = \av_{\rm r} \, . \label{3.9}
\end{equation}
Substituting (\ref{3.9}) into Eq.~(\ref{2.3}) gives
\begin{equation}
 \frac{\partial^2 \av}{\partial t^2} - c^2 \grad^2
\av = 4\pi c \jv_{\rm r} \, . \label{3.10}
\end{equation}
Eq.~(\ref{3.10}) indicates that in the Coulomb gauge (\ref{3.4})
the vector potential $\av$ propagates at speed $c$.

\section{Mechanical interpretation}

Recently, we have shown \cite{Dmitr} that in the Coulomb gauge
electrodynamics is isomorphic to the elastic medium that is stiff
to compression yet liable to shear deformations. In this analogy
the vector potential corresponds to the velocity and the scalar
potential to the pressure of the medium. Clearly, in an
incompressible medium there is no longitudinal waves, the pressure
acts instantaneously, and the transverse wave spreads at finite
velocity. This mechanical picture provides an intuitive support to
the electrodynamic relations (\ref{3.4}), (\ref{3.6}) and
(\ref{3.10}) just obtained.

\section{Conclusion}
By using a two-parameter Lorenz-like gauge, we extended
electrodynamics to a two-speed theory. Turning the longitudinal
speed parameter to infinity we come to electrodynamics in the
Coulomb gauge. In this way we show that the scalar potential acts
instantaneously while the vector potential propagates at speed of
light.

\begin{acknowledgements}
I would like to express my gratitude to Dr.\ I.\ P.\ Makarchenko
 for valuable comments concerning the
non-existence of longitudinal waves of the electric field and
longitudinal waves in the Coulomb gauge.
\end{acknowledgements}

\end{document}